# Symmetry breaking from Lorentz transformation

Bin-Guang Ma
Email: bgMa@sdut.edu.cn

**Abstract**

Symmetry breaking is discussed in this paper which comes from the Lorentz transformation of special relativity and changes our view that two relatively moving bodies always have the relative speed equal in magnitude and opposite in direction. An appeal was made for the test of Lorentz transformation in a multimedia environment which has the fundamental significance that it may reveal the origin of symmetry breaking in many physical phenomena.

In 1905, Albert Einstein published his special relativity that changed our view about space and time where Lorentz transformation is the foundation of its formulism. In Lorentz transformation, an inertia frame independent velocity scale was introduced which served by the light speed in vacuum[1]. However, bodies are not always moving in vacuum. If two bodies are relatively moving in two different media, such as a fish swimming in a lake and a bird flying in the sky, what will happen?

As shown in Fig. 1, suppose water is resting relative to air, and take the water/air system as reference frame. There is a ball moving in water and another ball moving in air. Now we consider the relative speeds between ball 1 and ball 2.

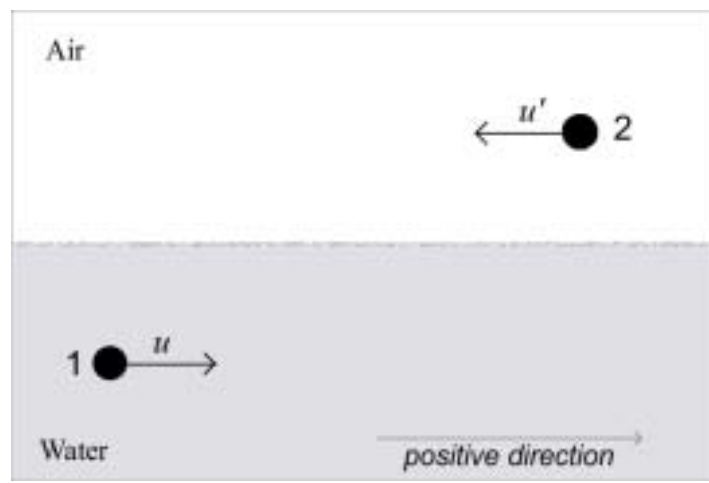

**Figure 1** Two relatively moving balls in two different media. The upper half is air and the lower half is water; air and water are resting relative to each other and taken as reference frame; ball 1 is moving towards right with velocity *u* relative to water; ball 2 is moving towards left with velocity *u'* relative to air. Right is assigned as the positive direction.

Firstly, we consider the velocity of ball 2 relative to ball 1. To measure the velocity of ball 2, ball 1 must receive the light from ball 2. Since ball 1 is moving at velocity *u* relative to water, equivalently, that is to say, water is moving at velocity –*u* relative to ball 1. Therefore, the velocity scale in Lorentz transformation should be the speed of the light from ball 2 measured by ball 1. According to Fresnel's drag formula, the speed of the light from ball 2 measured by ball 1 is

$$c_w = \frac{c}{n_w} + \left(1 - \frac{1}{n_w^2}\right)u \tag{1}$$

where $c$ is the light speed in vacuum and $n_w$ is the refractive index of water.

Considering the refractivity of medium, the velocity measured in water is not equal to that measured in air, there is

$$\frac{u'}{u'_w} = \frac{c^a}{c^w} = \frac{n_w}{n_a} \tag{2}$$

where $u'_w$ is the velocity of ball 2 if measured in water, $c^a$ is the light speed in air, $c^w$ is the light speed in water, $n_a$ is the refractive index of air. Thus,

$$u'_w = \frac{n_a}{n_w} u'. \tag{3}$$

According to Lorentz velocity addition formula, we get the velocity of ball 2 relative to ball 1 is

$$v_{12} = -\frac{u'_w + u}{1 + \frac{u'_w u}{c_w^2}}. \tag{4}$$

Substitute (1) and (3) into (4), we get

$$v_{12} = -\frac{u'_w + u}{1 + \frac{u'_w u}{c_w^2}} = -\frac{u + \frac{n_a}{n_w}u'}{1 + \frac{n_a}{n_w}\frac{u'u}{\left(\frac{c}{n_w} + \left(1 - \frac{1}{n_w^2}\right)u\right)^2}}. \qquad (5)$$

Secondly, we consider the velocity of ball 1 relative to ball 2. To measure the velocity of ball 1, ball 2 must receive the light from ball 1. Since ball 2 is moving at velocity *u'* relative to air, equivalently, that is to say, air is moving at velocity *–u'* relative to ball 2. Therefore, the velocity scale in Lorentz transformation should be the speed of the light from ball 1 measured by ball 2. According to Fresnel's drag formula, the speed of the light from ball 1 measured by ball 2 is

$$c_a = \frac{c}{n_a} + \left(1 - \frac{1}{n_a^2}\right)u'. \qquad (6)$$

Considering the refractivity of medium, the velocity measured in air is not equal to that measured in water, there is

$$\frac{u}{u_a} = \frac{c^w}{c^a} = \frac{n_a}{n_w} \qquad (7)$$

where $u_a$ is the velocity of ball 1 if measured in air. Thus,

$$u_a = \frac{n_w}{n_a}u. \qquad (8)$$

According to Lorentz velocity addition formula, we get the velocity of ball 1 relative to ball 2 is

$$v_{21} = \frac{u_a + u'}{1 + \frac{u_a u'}{c_a^2}}. \qquad (9)$$

Substitute (6) and (8) into (9), we get

$$v_{21} = \frac{u_a + u'}{1 + \frac{u_a u'}{c_a^2}} = \frac{u' + \frac{n_w}{n_a} u}{1 + \frac{n_w}{n_a} \frac{uu'}{\left(\frac{c}{n_a} + \left(1 - \frac{1}{n_a^2}\right) u'\right)^2}}. \tag{10}$$

Because $n_w \neq n_a$, therefore $v_{12} \neq -v_{21}$, i.e., two relatively moving bodies in two different media break the symmetry that two relatively moving bodies always have the relative speed equal in magnitude and opposite in direction. If the two balls have the same rest mass $m_0$, then according to Einstein's mass-energy relation, there are

$$E_{12} = \frac{m_0 c^2}{\sqrt{1 - \frac{v_{12}^2}{c^2}}} = \frac{m_0 c^2}{\sqrt{1 - \frac{\left(u + \frac{n_a}{n_w} u'\right)^2}{\left(1 + \frac{n_a}{n_w} \frac{u'u}{\left(\frac{c}{n_w} + \left(1 - \frac{1}{n_w^2}\right) u\right)^2}\right)^2 c^2}}} \tag{11}$$

and

$$E_{21} = \frac{m_0 c^2}{\sqrt{1 - \frac{v_{21}^2}{c^2}}} = \frac{m_0 c^2}{\sqrt{1 - \frac{\left(u' + \frac{n_w}{n_a} u\right)^2}{\left(1 + \frac{n_w}{n_a} \frac{uu'}{\left(\frac{c}{n_a} + \left(1 - \frac{1}{n_a^2}\right) u'\right)^2}\right)^2 c^2}}}. \tag{12}$$

The former is the energy of ball 2 from the viewpoint of ball 1, and the latter is the energy of ball 1 from the viewpoint of ball 2. Because $n_w \neq n_a$, there is an energy difference between them that

$$\Delta E = |E_{12} - E_{21}|. \tag{13}$$

Suppose $m_0 = 1$ kg , $c = 299792458$ m/s , $n_a = 1.003$ , $n_w = 1.333$ , $u = 5$ m/s , $u' = 8$ m/s , then

$$\Delta E = | E_{12} - E_{21} | = 46.524 \text{ J} \tag{14}$$

Now the question is: Is it possible that an experiment is designed to test this difference of energy? This will be a new test of Lorentz transformation in a multimedia environment. Whatever the result is, it will lead to a new comprehension of Lorentz transformation.

Someone may argue that the velocity scale in Lorentz velocity-addition formula (namely, formula (4) and (9)) should also be the light speed in vacuum. Even so, the symmetry breaking still exists. If we replace $c_w$ in (4) with $c$, we get

$$v_{12} = -\frac{u'_w + u}{1 + \dfrac{u'_w u}{c^2}} = -\frac{\dfrac{n_a}{n_w} u' + u}{1 + \dfrac{n_a}{n_w} \dfrac{u'u}{c^2}} . \tag{15}$$

Similarly, if we replace $c_a$ in (9) with $c$, we get

$$v_{21} = \frac{u_a + u'}{1 + \dfrac{u_a u'}{c^2}} = \frac{\dfrac{n_w}{n_a} u + u'}{1 + \dfrac{n_w}{n_a} \dfrac{uu'}{c^2}} . \tag{16}$$

Then, according to Einstein's mass-energy relation, we get

$$E_{12} = \frac{m_0 c^2}{\sqrt{1 - \dfrac{v_{12}^2}{c^2}}} = \frac{m_0 c^2}{\sqrt{1 - \dfrac{\left(\dfrac{n_a}{n_w} u' + u\right)^2}{\left(1 + \dfrac{n_a}{n_w} \dfrac{u'u}{c^2}\right)^2 c^2}}} \tag{17}$$

and

$$E_{21} = \frac{m_0 c^2}{\sqrt{1 - \frac{v_{21}^2}{c^2}}} = \frac{m_0 c^2}{\sqrt{1 - \frac{\left(\frac{n_w}{n_a} u + u'\right)^2}{\left(1 + \frac{n_w}{n_a} \frac{uu'}{c^2}\right) c^2}}} . \quad (18)$$

Because $n_w \neq n_a$, there is an energy difference between them that

$$\Delta E = | E_{12} - E_{21} | \neq 0 . \quad (19)$$

Suppose $m_0 = 1$ kg , $c = 299792458$ m/s , $n_a = 1.003$ , $n_w = 1.333$ , $u = 5$ m/s , $u' = 8$ m/s , then

$$\Delta E = | E_{12} - E_{21} | = 46.524 \text{ J} . \quad (20)$$

Although formula (11) and (12) are different from (17) and (18), respectively, the resultant $\Delta E$ ( = |(11)-(12)| or |(17)-(18)| ) only has very slight difference: $3.08 \times 10^{-14}$ J. This result may mean that symmetry breaking mainly originates from the refractivity between media rather than Lorentz transformation. But a careful thinking will reach that the refractivity between media and the different velocity scales in Lorentz transformation amount to the same thing in that the refractivity between media just originates from the difference of light speeds in them.

The symmetry breaking from Lorentz transformation has fundamental significance that it may be the origin of symmetry loss we are looking for in our fundamental physics[2]. Since matter consists of different kinds of particles – gauge bosons, fermions (leptons and quarks), gravitons, or even Higgs particles – and these particles never rest and their relative motions keep to Lorentz transformation (as what we know by far), the heterogeneity of matter or vacuum (such as vacuum polarization or so) will break symmetry spontaneously due to the variance of velocity scale in

Lorentz transformation, which means that symmetry breaking may originate from particles' environments and their mutual relations, i.e., symmetry breaking may be a systematic emergence.

Recent works by Gupta have revealed that gravity may be the Second-Order Relativistic-Manifestation of Electrostatic-Force[3]. Based on the relativistic velocity addition formula, Gupta deduced a net attractive force between two bodies composed of atoms, which supports the possibility that symmetry breaking may occur spontaneously as a systematic emergence due to Lorentz transformation.

**Acknowledgement** Many thanks to Prof. Gupta for helpful discussions. This work was partially supported by National Key Project for Basic Research (2003CB114400) and National Natural Science Foundation of China (Grant No. 30100035).

**Affiliation:**

[1] Shandong Provincial Research Center for Bioinformatic Engineering and Technique,

Center for Advanced Study, Shandong University of Technology,

Zibo 255049, P. R. China

Telephone: ++86-533-2780271; Fax: ++86-533-2780271;

[2] Departments of Physics, Tianjin University, Tianjin, 300072, China